\documentclass[aps,showpacs,prl,twocolumn,footinbib]{revtex4-1}

\usepackage{epsfig}
\usepackage{amsfonts}
\usepackage{amsmath}

\usepackage[utf8]{inputenc}
\usepackage[colorlinks,bookmarks=false,citecolor=blue,linkcolor=blue,urlcolor=blue]{hyperref}

\begin{document}

\title{Fractional Wigner crystal in the helical Luttinger liquid}
\author{N. Traverso Ziani$^{1}$, F. Cr\'epin$^{1}$ and B. Trauzettel$^{1}$}
 \affiliation{$^1$ Institute for Theoretical Physics and Astrophysics, University of W\"{u}rzburg, 97074 W\"{u}rzburg, Germany\\
 }
\date{\today}

\begin{abstract}
The properties of the strongly interacting edge states of two dimensional topological insulators in the presence of two particle backscattering are investigated. We find an anomalous behavior of the density-density correlation functions, which show oscillations that are neither of Friedel nor of Wigner type: they instead represent a Wigner crystal of fermions of fractional charge $e/2$, with $e$ the electron charge. By studying the Fermi operator, we show that the state characterized by such fractional oscillations still bears the signatures of spin momentum locking. Finally, we compare the spin-spin correlation functions and the density-density correlation functions to argue that the fractional Wigner crystal is characterized by a non trivial spin texture.
\end{abstract}

\pacs{71.10.Pm., 73.22.Lp, 73.21.-b}
\maketitle
The helical Luttinger liquid\cite{helical1} (hLL) is the state of electronic matter that describes the interacting helical edges of the recently predicted\cite{bernevig,km1,km2} and experimentally realized\cite{koenig7,altro} two dimensional topological insulators\cite{hasan,qi,moore} (2DTI). Particular emphasis has been devoted to the investigation of the transport properties of 2DTI: topological protection of the edge states facilitates the observation of conductance quantization\cite{koenig7,altro,nonlocal} in short samples. Long edges on the other hand are characterized by a reduced conductance. Even though a comprehensive theoretical understanding of the scattering sources causing the reduction of the conductance of the edge is still lacking, the role of electron-phonon interactions\cite{phonons}, of magnetic\cite{magnetic} and nonmagnetic impurities\cite{helical2,add1}, in the presence of random Rashba disorder\cite{new,add2}, of breaking of axial symmetry in combination with electron-electron interactions and impurities\cite{generalhelical,add3}, of tunneling among the edges and charge puddles in the bulk of the 2DTI\cite{puddles}, and of the coupling between opposite edges\cite{finite1,giacomo} has been theoretically elucidated. The mathematical tool allowing for most of such calculations is bosonization\cite{giamarchi,voit}, a procedure that enables us to recast the Hamiltonian of the interacting electrons on the edges into a Hamiltonian of free bosonic excitations, representing charge density waves, and to express the Fermi operator in terms of the creation and annihilation bosonic operators. The physical meaning of the bosonization technique can be understood within the framework of Luttinger liquid theory\cite{haldane}, that is the one dimensional counterpart of the Fermi liquid theory for one dimensional gapless systems. More precisely, the exactly solvable Luttinger model\cite{luttingermodel,vondelft,libronelieb}, a strictly linear theory of interacting one dimensional fermions with infinite bandwidth in the single particle dispersion is usually employed. The validity of the Luttinger model as a basis for the description of interacting electrons has a number of experimental demonstrations, ranging from spin charge separation\cite{spincharge}, to charge fractionalization\cite{fractionalization1,fractionalization2}, and to anomalous tunneling\cite{anomaloust1,anomaloust2}. On the other hand, the Luttinger model alone fails in predicting a reasonable behavior of local observables\cite{eggert,num1,noi,safi}, such as the electron density and the density-density correlation functions when electron-electron interactions are strong. In particular, the Luttinger model is not able to capture the transition between a weakly correlated state dominated by Friedel oscillations of the density\cite{open,haldanelong}, and the strongly correlated one dimenional Wigner crystal\cite{wigner,nat1,nat2,fiete,schulz,mfg1,mfg2,vignale}. To overcome this problems one has to consider a richer theory: the Luttinger {\it liquid}, the universal model describing low energy properties of gapless one dimensional systems\cite{giamarchi,voit,haldane}. The construction by Haldane\cite{haldane} clearly shows that the Fermi operator $\psi_s(x)$, where $s=\pm$ is the spin projection, of a generic one dimensional electron system can be significantly different with respect to the one of the Luttinger model\cite{haldane,safi}: the standard relation $\psi_s(x)\sim e^{i\theta(x)}\sum_{p=\pm1}e^{-ipk_Fx}e^{ip\phi(x)}$, with $k_F$ the Fermi momentum and $\phi(x)$ and $\theta(x)$ the usual bosonic fields, is replaced by the more general expression $\psi_s\sim e^{i\theta(x)}\sum_{p=-\infty}^{\infty}c^{(s)}_p e^{-ipk_Fx}e^{ip\phi(x)}$, with the model dependent coefficients $c^{(s)}_p$.

We aim at understanding the properties of the strongly interacting hLL, and hence we have to build the appropriate Luttinger liquid theory. This formulation presents difficulties since some of the usual paradigms break down:

(\textit{i}) spin momentum locking breaks the symmetry, usually holding for standard Luttinger liquids, $c^{(s)}_{-p}=c^{(s)}_p$, since it would imply no preferred chirality for a given spin projection,

(\textit{ii}) time reversal symmetry protects from one particle backscattering off nonmagnetic impurities, which implies that the usual Friedel oscillations of the density are forbidden: in fact, if they were present a capacitive coupling between the electrons of the edge and the impurity potential would lead precisely to one particle backscattering.

In this Letter, we develop a Luttinger liquid picture of the strongly interacting quantum spin Hall system in the presence of two-particle backscattering extending over all the helical edge. We find a state characterized by charge oscillations. These oscillations are profoundly different from usual Friedel or Wigner ones in view of their different wavelength: they are characterized by a wavelength that is half of the wavelength of the usual Wigner crystal, suggesting the formation of a correlated state of fermions with charge $e/2$, with $e$ the electron charge. Moreover we show, by studying the series expansion of the Fermi operator, that peculiar features, inherited by spin momentum locking, still survive in the strongly interacting regime characterized by density oscillations. Finally we address the spin-spin correlations and demonstrate that they have the wavelength of the usual Wigner oscillations, suggesting a complex spin pattern.\\

The Hamiltonian of the system in its bosonized form can be written as
\begin{equation}
H=\int_0^L dx \left\{\left[\frac{\prod}{2}+\frac{\left(\partial_x\phi\right)^2}{2}\right]-\frac{\mu\beta}{4\pi}\partial_x\phi+h_{2p}\right\}\label{eq:H}
\end{equation}
where the first contribution is the usual Luttinger liquid Hamiltonian, the second is the chemical potential term, and $h_{2p}$, given by
\begin{equation}
h_{2p}=\frac{g_{2p}}{2(\pi \alpha)^2}\cos(\beta  \phi),
\end{equation}
is the two particle backscattering. More specifically $\phi$ is the Luttinger liquid bosonic field, $\prod$ its conjugated field, $\mu$ is the chemical potential, $\beta$ is related to the Luttinger parameter $K_L$ by $\beta=4\sqrt\pi K_L$, $g_{2p}$ measures the strength of the two-particle backscattering and $\alpha$ is the cutoff of the Luttinger theory. Periodic boundary conditions, with period $L$, are imposed. Two-particle backscattering can arise in the presence of anisotropic spin interactions\cite{helical1} or in generic helical liquids\cite{orth}; it is allowed by time reversal symmetry and its effect is more pronounced when the Fermi level is close to the Dirac point. Our interest in such an interaction term is due to its formal analogy with the umklapp term occurring in usual Luttinger liquids, which is known to lead to Wigner oscillations in the density\cite{giamarchi,eggert}, and finally to the formation of Wigner molecules\cite{physicaE} in finite systems.

We are interested in analyzing the helical counterpart of the Wigner crystal, and hence we focus on the regime of strong interaction $\beta\ll 1$. An effective strategy to deal with such a regime in the case of a regular spinful Luttinger liquid has been developed in Ref.\onlinecite{aristov}. The same strategy can also be applied to the case of the hLL. The first step, after defining for notational convenience the mass $m=\sqrt{g_{2p}\beta^2}/(2\pi^2a^2)$ of the so-called breather, is to recast the problem into the Lagrangian formalism. The Lagrangian density $\mathcal{L}$ reads
\begin{equation}
\mathcal{L}=\frac{\left(\partial_t\phi\right)^2}{2}-\frac{\left(\partial_x\phi\right)^2}{2}-\frac{m^2}{\beta^2}\cos(\beta  \phi)+\frac{\mu\beta}{4\pi}\partial_x\phi.
\end{equation}
The classical solution  $\phi_0$ of the equation of motion, upon which a low energy theory for fluctuations can be obtained with standard techniques, becomes\cite{collective}
\begin{equation}
\phi_0(x,X)=\frac{2}{\beta}\mathrm{am}\left[ \frac{m(x+X)}{k},k \right]\label{eq:class}
\end{equation}
with $\mathrm{am}\left[ y,k \right]$ the Jacobi amplitude function with elliptic index $k$. The quantity $X\in[0,L]$ is a free parameter whose presence is due to translational invariance. Physically it represents the 'center of mass' of the soliton solution. The index $k$ is fixed by the the number $Q$ of electrons on the helical edge, as measured from the Dirac point: since the density $\rho_0(x)$, corresponding to the classical solution, is given by
\begin{equation}
\rho_0(x,X)=\frac{\beta}{4\pi}\partial_x\phi_0(x,X),\label{eq:classdens}
\end{equation}
one has
\begin{equation}
\frac{mL}{4Q}=kK(k)\label{eq:central},
\end{equation}
with $K(k)$ the complete elliptic integral of the first kind. In fact, by integrating Eq.(\ref{eq:classdens}), we obtained $\phi(L-X,X)-\phi(-X,X)=4\pi Q/\beta$; the same difference can be computed using Eq.(\ref{eq:class}). By matching the two expressions one obtains the result in Eq.(\ref{eq:central}).

\begin{figure}[htbp]
\begin{center}
\includegraphics[width=8cm,keepaspectratio]{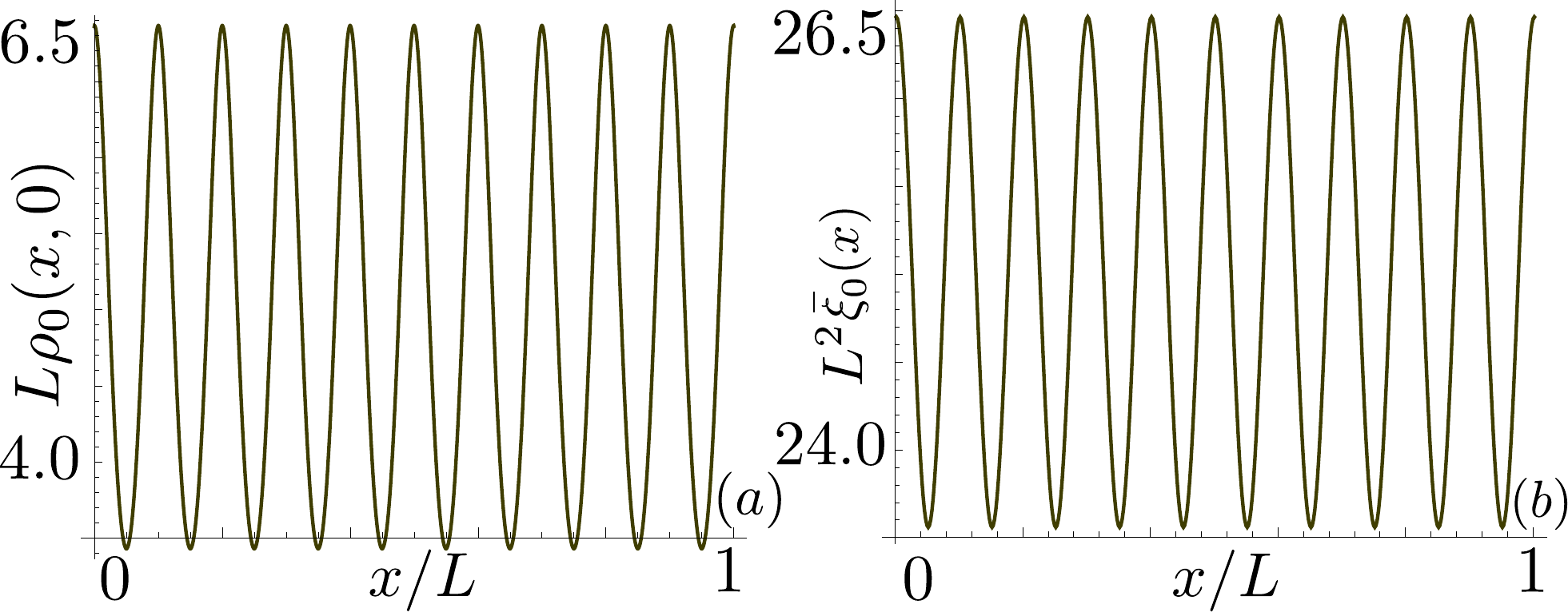}
\caption{(Color online) Classical electron density $\rho_0(x,0)$ (panel (a)) and density-density correlation function $\bar\xi(x)$ (panel (b)), for Q=5 and $Q/mL=0.15$.}
\label{fig:classical}
\end{center}
\end{figure}

The number of particles $Q$ is instead controlled by the chemical potential, via the minimization of the total energy. The condition expressed in Eq.(\ref{eq:central}), which is different from the one occurring in usual Luttinger liquids\cite{aristov}, has important implications on physical observables. Although a deeper analysis, which will be discussed below, requires the inclusion of quantum fluctuations, a flavour of the physical consequences of two-particle backscattering can be gained already by considering the properties of the classical solution. We can consider the electron density $\rho_0(x,X)$ once a particular choice of $X$, say $X=0$, is done. This quantity describes the local electron density if a local perturbation able to pin it is introduced. It is important to note, however, that when the average (integration) over $X$ is carried out, as required for the clean system described in Eq.(\ref{eq:H}), the average local density $\bar{\rho}_0=Q/L$ is constant due to translational invariance. Alternatively we can consider the density-density correlation function $\bar{\xi}(x)$ in the classical regime, after averaging (integrating) over the variable $X$. The average over $X$ is physically needed since the center of mass of the classical solution can be anywhere in the ring with equal probability. Explicitly one has
\begin{eqnarray}
\rho_0(x,0)&=&\frac{2QK(k)}{\pi L}\mathrm{dn}\left[\frac{4QK(k)x}{L},k\right],\\
\bar{\xi}_0(x)&=&\frac{1}{L}\int_0^L dX \rho_0(x,X)\rho_0(0,X),
\end{eqnarray}
where $\mathrm{dn}\left[x,k\right]$ is the Jacobi dn function with elliptic index $k$. As shown in Fig.\ref{fig:classical}, the number of peaks of $\rho_0(x,0)$ and $\bar{\xi}(x)$ is $2Q$, instead of $Q$, as it is in the usual Luttinger liquid, and as would be expected if the system was in the Wigner molecule regime. It will be soon shown that this feature is also present when quantum fluctuations are included, and is hence a characteristic of the helical liquid: the doubling of the number of peaks in the strong interaction regime is due do the fact that time reversal symmetry does not allow for one particle backscattering: one particle backscattering would lead to the expected $Q$ peaks in $\rho_0(x,0)$ and $\bar{\xi}(x)$.
The scenario can be interpreted as a Wigner oscillation of quasi-particles with charge $e/2$: in the very same way strong interaction favors Wigner oscillations of electrons with integer charge in usual spinful Luttinger liquids, here the interplay of spin momentum locking, two-particle backscattering, and electron-electron interactions, lead to the formation of such fractional oscillations. A physical insight in the meaning of this Wigner oscillation of fractional charges can be gained by refermionization: this technique allows to exactly solve the Hamiltonian in Eq.(\ref{eq:H}) for $K_L=1/4$, and for that value of the interaction parameter one has half charged fermions as low energy excitations. Since two-particle backscattering becomes relevant in the RG sense for $K_L<1/2$, and we are considering $K_L\ll 1$, the quasiparticles of our system are also expected to be fermions with charge $e/2$.

In order to refine the classical result, a theory for quantum fluctuations $\eta(x)$ must be addressed. The appropriate framework is the collective coordinate method\cite{collective}, and the most appropriate way of including fluctuations is to introduce them by the relation\cite{aristov}
\begin{equation}
\phi(x)=\frac{2}{\beta}\mathrm{am}\left[ 2K(k)\left(\frac{2Qx}{L}+\frac{\beta\eta(x)}{2\pi}\right),k \right],\label{eq:fluct}
\end{equation}
which reduces to Eq.(\ref{eq:class}) for $\eta(x)=0$, up to the variable $X$. This behavior suggests the existence of a zero energy mode $\eta_0(x)$ which is just a constant and stems for translational invariance. The expression for the electron density $\rho(x)$ resulting from Eq.(\ref{eq:fluct}) can be conveniently written by using a standard series expansion for the Jacobi dn function as\cite{special}
\begin{equation}
\rho(x)=\left(\frac{Q}{L}+\frac{\beta}{4\pi}\partial_x\eta(x)\right)\sum_{n=-\infty}^\infty\frac{e^{in(4\pi Qx/L+\beta\eta)}}{\cosh(n\tau)}\label{eq:densitaserie}
\end{equation}
with $\tau=\pi K(\sqrt{1-k^2})/K(k)$. The expression in Eq.(\ref{eq:densitaserie}) closely resembles Haldane's expansion for the electron density in the spinless Luttinger liquid\cite{haldane}, though with a striking difference: the harmonics of the density appear with wavevectors which are multiples of $4k_F=4\pi Qx/L$, instead of multiples of $2k_F$. This behavior is consistent with the oscillations in the density corresponding to the classical solution (Fig.\ref{fig:classical}). Note that both in the limit $m\rightarrow 0$ (no cosine term) and in the limit $Q/(LM)\rightarrow \infty$ (significantly away from the Dirac point, that is where the effects of two-particle backscattering are expected to be negligible) only the term with $n=0$ is present due to the presence of the damping factors $\cosh(n\tau)$. The usual form of the electron density of the hLL, i.e. the long wave density of usual one-channel Luttinger liquids, is hence recovered in such limits, as expected.\\
 The energetics of the fluctuations $\eta(x)$ is encoded, up to the quadratic order in $\eta(x)$, in the Lagrangian density\cite{aristov}
\begin{equation}
\mathcal{L}_\eta(x)=\frac{w^2(x)}{2}\left[(\partial_t\eta)^2-(\partial_x\eta)^2\right]\label{eq:fluctL}
\end{equation}
with
\begin{equation}
w(x)=\frac{4K(k)}{\pi}\mathrm{dn}\left[\frac{4QK(k)x}{L},k\right].
\end{equation}
We could exactly solve the equation of motion, deriving from Eq.(\ref{eq:fluctL}), for the eigenmodes $\eta(x,t)=\mathrm{exp}[-i\omega_j t]\eta_j(x)$. In fact this equation becomes
\begin{equation}
\omega_j^2\eta_i(x)=w^{-2}(x)\partial_x\left(w^2(x)\partial_x\eta_j(x)\right),
\end{equation}
and it maps onto a Lam\'e equation. The lowest energy eigenfunction, which corresponds to $\omega=0$, is $\eta_0(x)=\mathrm{const.}$, as required by the translation invariance of the system. When quantizing the theory this mode can be treated within the collective quantization approach, even though in the large $Q$ limit it has been shown to be sufficient to integrate over $\eta_0$\cite{aristov}. At low energy, still in the large $Q$ limit, the theory for the low energy modes $\eta_j(x)$, with $j\neq 0$, is essentially equivalent to a one channel Luttinger liquid\cite{aristov}, with Fermi velocity $v_F=\sqrt{1-k^2}K(k)/E(k)$ and Luttinger parameter $K'_L=\pi\beta^2/(16\sqrt{1-k^2}K^2(k))$.\\
By virtue of Eqs.(\ref{eq:densitaserie}) and (\ref{eq:fluctL}) in its low energy limit, we could show that the average electron density is, as expected, constant, and the zero temperature density-density correlation function $\bar\xi(x)$ is, in the limit $L\rightarrow\infty$ and up to the slowest decaying oscillating term,
\begin{equation}
\bar\xi(x)=\frac{Q^2}{L^2}+\frac{K'_L}{2\pi^2x^2}+\bar\xi_{4k_F}(x)
\end{equation}
with
$\bar\xi_{4k_F}(x)\sim {\cos\left(\frac{4\pi Qx}{L}\right) }/{x^{2K'_L}}.$
The absence of the $2k_F$ component, which is the one that indicates the formation of a Wigner crystal, or at least Wigner oscillations, in one channel Luttinger liquids, witnesses that the strongly interacting sector of the hLL is profoundly different from the usual strongly interacting electron gas, and that the hLL does not undergo a transition to a regular one dimensional Wigner crystal. In fact two particle backscattering leads to a state characterized by charge fractionalization and the emergent quasiparticles (with fractional charge $e/2$) are hence $2Q$ due to charge conservation. They interact among each other to form the fractional Wigner oscillations.

In order to deepen the understanding of the effects of the original helical character of the theory on the strong interaction regime, it is natural to investigate how much spin momentum locking is affected when strong two-particle backscattering is included. A natural way to gain insight into this topic is to address the form of the spinor representing the Fermi operator. To do so one has to build the field $\theta(x)$, conjugated to $\eta(x)$. Leaving aside the discussion of the zero energy component, that is not relevant for the following, it is easy to show\cite{aristov} that the field
$\theta(x)=\int^x\tilde{\prod}(x')dx',$
with $\tilde{\prod}(x)=\partial\mathcal{L}/\partial(\partial_t\eta)$ and with the quantization condition $[\eta(x),\tilde{\prod}(x')]=i\delta(x-x')$, can be evoked to properly define the Fermi operator for right/left moving electrons. By using the series expansion
$e^{i\mathrm{am}[2K(k)z]}=\sum_{n=-\infty}^\infty C_n e^{2\pi iz(n+1/2)}$,
with $C_n={\pi e^{\tau(n+1/2)}}/({kK(k)\sinh(2n+1)\tau})$, and the definition of the chiral fields  $\phi_{R/L}(x)=\frac{\beta}{4}\phi(x)\mp\frac{2\pi}{\beta}\theta(x)$
we obtain the expression for the Fermi fields in terms of $\eta(x)$ and $\theta(x)$
\begin{equation}
\psi_{R/L}(x)\propto e^{-2i\pi\beta^{-1}\theta(x)}\sum_{n=-\infty}^\infty C^{(1)}_n e^{\pm i (4\pi Qx/L+\beta\eta)(n+1/4)},\label{eq:fermioperator1}
\end{equation}
where $C_n^{(1)}$ is implicitly given by the relation
$C_n^{}=\sum_{n=-\infty}^\infty C_{n-n_1}^{(1)}C_{n_1}^{(1)}$.
The form of the Fermi spinor $\Psi(x)$ is hence $\Psi(x)=(\psi_+(x),\psi_-(x))^T=(\psi_R(x),\psi_L(x))^T,$
where $\psi_{\pm}(x)$ is the spin up/down component of the field operator. The usual form of the hLL can be recovered by setting $C_n^{(1)}\propto \delta_{n,0}$, as can be easily shown to be the case in the limits $m\rightarrow 0$ and $Q/(mL)\rightarrow\infty$. For finite $m$ and $Q/(mL)$ the breaking of spin momentum locking is evident: the terms with $n>0$ ($n<0$) represent left (right) moving components in the spin up (down) part of the spinor. However these terms are suppressed by both the coefficients $C_n^{(1)}$ and the scaling they acquire in the correlation functions. Since it is a direct consequence of the chirality of the Fermi spinor, the strong anisotropy in the spin response function, hallmark of the hLL, is hence to some extent also present in its strong interaction sector in the presence of two-particle backscattering. This is in sharp contrast with the behavior of the one dimensional Wigner crystal, which is an almost classical state in which spin dynamics plays an unimportant role, and is often integrated out\cite{fiete}.
To support our claim we address the ground state average $s_{ij}(x)$ of the spin-spin functions $S_{ij}(x)$ given by $S_{ij}(x)=\Psi^\dag(x)\sigma^i\Psi(x)\Psi^\dag(0)\sigma^j\Psi(0)$.\\

\begin{figure}[htbp]
\begin{center}
\includegraphics[width=7.cm,keepaspectratio]{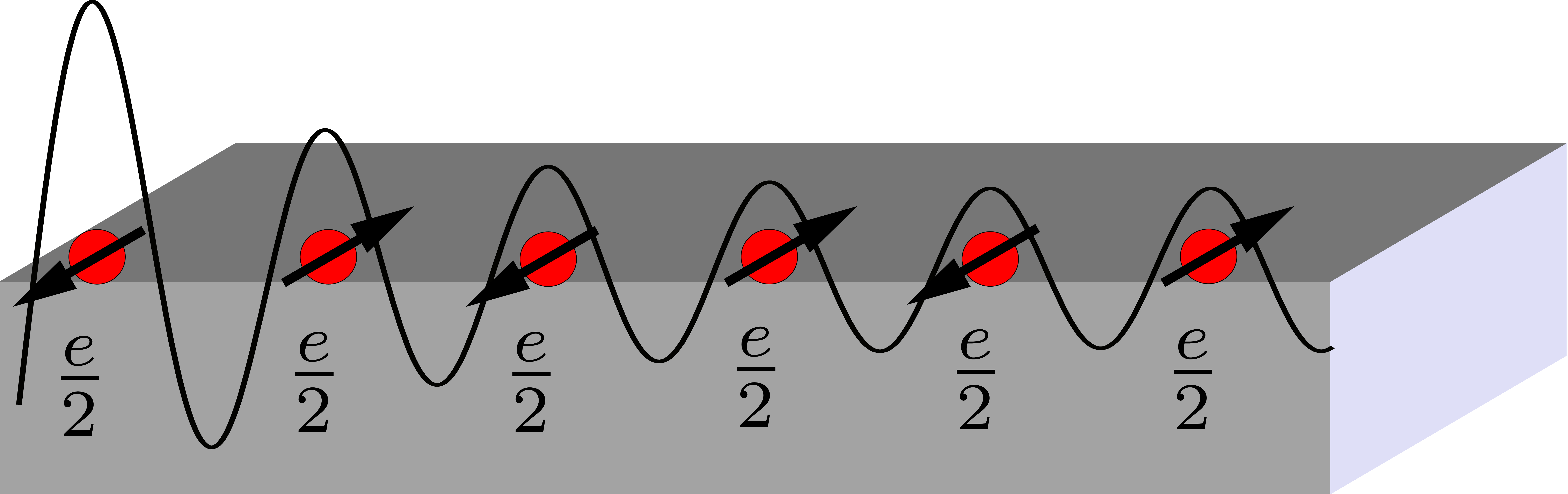}
\caption{(Color online) Scheme of the strongly interacting ground state: the fractional quasiparticles tend to form a Wigner crystal (the density-density correlation function is shown). Moreover the spin density is schematically depicted by  using the arrows.}
\label{fig:2}
\end{center}
\end{figure}

In the hLL without two-particle backscattering one finds that $s_{11}(x)=s_{22}(x)$ and $s_{12}(x)=-s_{21}(x)$ are purely oscillating functions with wavevector $2k_F$, while $s_{33}(x)$ is a non oscillating function and the remaining correlations vanish. In our case, $s_{33}(x)$ is still a non oscillating function, so that the probability of finding two electrons at distance $x\gg \alpha$ from each other is independent of the spin projection and follows the density-density correlation function. On the other hand the oscillations in $s_{11}(x)$, $s_{22}(x)$ and $s_{12}(x)$ have a wavelength which is twice the wavelength of the oscillations of the electron density and hence they witness the onset of a spin helix: this behavior is in accordance with what happens in the weakly interacting hLL. However the spin oscillations in the strongly interacting edge with two particle backscattering are built on oscillating density-density correlations characterized by half of the wavelength, suggesting an intercalation of electrons with spin up and spin down, once a projection axis perpendicular to the $z$ axis is chosen. An intuitive picture is drawn in Fig.\ref{fig:2}.

In conclusion we have shown that two-particle backscattering, in combination with strong electron electron interaction, leads to the formation of a peculiar correlated structure, a Wigner crystal of quasi-particles with charge $e/2$. This state still bares the signatures of the underlying helicity of the system, both in the form of the Fermi operator and in the anisotropy of the spin-spin correlation functions. Moreover such a state has a complex spin structure, resembling a spin helix, that suggests an intercalation of particles with opposite spin projection once a projection axis perpendicular to the $z$ axis is chosen.

We acknowledge financial support by the DFG (German-Japanese research group on "Topotronics" and SPPI666), the Helmholtz Foundation (VITI), and the ENB Graduate school on "Topological Insulators". We thank R. Fazio for interesting discussions.

\end{document}